# Siamese Residual Neural Network for Musical Shape Evaluation in Piano Performance Assessment

Xiaoquan Li[1], Stephan Weiss[1], Yijun Yan[2], Yinhe Li[2], Jinchang Ren[2], John Soraghan[1], Ming Gong[1]
[1]*Centre for Signal and Image Processing, Dept. Electronic & Electrical Engineering, University of Strathclyde, Glasgow, UK*
[2]*National Subsea Centre, School of Computing, Robert Gordon University, Aberdeen, UK*
Email: {xiaoquan.li, stephan.weiss, j.soraghan, ming.gong}@strath.ac.uk; {y.yan2, y.li24, j.ren}@rgu.ac.uk;

*Abstract*— Understanding and identifying musical shape plays an important role in music education and performance assessment. To simplify the otherwise time- and cost-intensive musical shape evaluation, in this paper we explore how artificial intelligence (AI) driven models can be applied. Considering musical shape evaluation as a classification problem, a light-weight Siamese residual neural network (S-ResNN) is proposed to automatically identify musical shapes. To assess the proposed approach in the context of piano musical shape evaluation, we have generated a new dataset, containing 4116 music pieces derived by 147 piano preparatory exercises and performed in 28 categories of musical shapes. The experimental results show that the S-ResNN significantly outperforms a number of benchmark methods in terms of the precision, recall and F1 score.

## I. INTRODUCTION

As a sub-task of music information retrieval (MIR), music performance assessment (MPA) has drawn considerable attention [1, 2]. In practical music education, MPA helps to improve a candidate's ability to perform. However, it needs comprehensive music perception and cognition which are mostly built up via long-term practice and understanding. Currently, MPA for piano students is mostly guided by music trainers, resulting in extensive time and resources needed. An AI-driven MPA would be particularly useful for raising the efficacy of music education whilst reducing the cost (Fig. 1).

The marking criteria of music scores in some music educational exams, such as the Associated Board of the Royal Schools of Music (ABRSM) in the UK, are determined by five elements, i.e., pitch, time, tone, shape and overall performance [3]. Many machine learning and deep learning models have been explored to evaluate the pitch, time, rhythm and tone in MPA [2]. Typical works include convolutional neural network (CNN) for local tempo and tempo stability assessment [4], support vector regressor for assessing rhythmic accuracy and tone quality [5], fusion of CNN and recurrent network for pitch extraction [6], and the integration of 2DCNN and 3DCNN for evaluation of piano skills [7], etc.

However, musical shape evaluation for MPA remains unaddressed to date. Musical shape, as a unit to build a coherent narrative music environment, is one of the most evident characteristics, which offers a more ecologically valid way of understanding the feeling responses to the music than seeing music as expressive of particular emotional states [8]. Due to its abstraction and multisensory perception, the musicology study has progressed slowly. In 1963, Langer [9] proposed a famous theory of 'sound the way moods feel' where musical shape is considered as a fundamental unit in music's intrinsic properties. In 2006, Eitan and Granot [10] discussed the associations of dynamics, pitch, time, and articulation to musical shape by comparing the music perception with musicians and non-musicians. The result found that time patterns are highly related to the musical shapes. Pitch and dynamics patterns are potentially linked with musical shape, which required further investigation. To tackle this issue, Küssner and Leech-Wikinson [11] carried out an extensive study in 2014 and discovered that pitch does not closely associate with musical shape. In 2017, Daniel [8] defined that musical shape refers to the small dynamics changes in music that can represent feeling and movement states, or any tiny changes varying with time, giving life-like qualities to music. Therefore, the core in musical shape evaluation is to discriminate the time and dynamics patterns in music pieces.

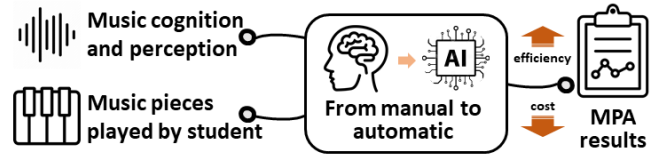

Fig. 1. Concept of current MPA in music education.

Lerch presented a critical review of MPA [12] and pointed out that most MIR researchers neglect the difference between score-like and performance information. Although many scholars have studied time and dynamics patterns with the corresponding datasets reported [12], they do not intend to evaluate the music's intrinsic properties. Thus, it brings a large barrier for using existing MIR technique to interpret the music's intrinsic properties. To break through this bottleneck, it is crucial to add a new dimension to MPA, conducting the study of musical shape evaluation to align the human perception with music's intrinsic properties. For this purpose, a specific dataset is also needed. Thus, we report on a new musical shape evaluation dataset (see details in Section 2) where all music pieces are performed by 3 music trainers and 10 young students. Inspired by the concept shown in Fig. 1, we define the music pieces performed by music trainers based on their music cognition and perception having the normal musical shape and the corresponding ones performed by students having the specific musical shape to be identified.

From a signal processing perspective, the problem we are addressing is the evaluation of musical shape based on the integration of two key typical parameters: dynamics and tempo which capture two important aspects of a musical performance, i.e., the level of loudness and the speed of the music, respectively. For this purpose, some typical audio feature extraction methods have been explored [13]. Then machine learning model such as Support Vector Machine (SVM) can be used for improved decision making [14, 15]. On a different note, musical shape is an example of how multisensory capacity is facilitated by the sensorimotor cortex in the brain [8], and the neocortex is organized in a manner to make the underlying processes as efficient as possible. This has motivated us to develop a deep neural model for musical shape evaluation. Currently, many variant deep learning methods have been explored based on VGG16 [16], ResNet50 [17] and DenseNet161 [18] for audio feature extraction. However, most of them are not designed for music education.

A Siamese network is an architecture with two identical branches, where each takes one set of input data and the weights and bias of any neural network in each branch are the same [19]. The advantage of this architecture is that it can learn semantic similarity from the two inputs and does not need to rely on large datasets to perform well [20]. On the other hand, residual blocks [17] have attracted much attention thanks to the strong feature representation capability. As a result, its variants have been widely used for image classification and can produce impressive performance [21].

Taking the advance of both deep learning frameworks and following the concept shown in Fig. 1, we propose a Siamese residual neural network (S-ResNN) in this paper. In this model, the music piece played by music trainer and the corresponding one played by student are taken as the inputs which will be transformed into spectrum before sending to both branches. Then the proposed S-ResNN will extract the global spectral feature and identify the musical shapes in the piano pieces. Extensive experiments on our proposed musical shape evaluation dataset have shown the superiority of S-ResNN when comparing with the combination of machine learning and conventional signal processing methods, and three typical deep learning models. The main contributions of this paper are highlighted below.

- We explore a new dimension of MPA, which connects the human perception with music intrinsic properties;

TABLE I. DESCRIPTION OF 8 BASIC SHAPES.

| No. | Shape | | Description |
|---|---|---|---|
| 1 | *Forte* | Dynamics | Strong dynamics and denoted as *f* on the score |
| 2 | *Piano* | | Weak dynamics and denoted as *p* on the score |
| 3 | *Cresc.* | | Gradually increase the dynamics from *p* to *f* |
| 4 | *Decresc.* | | Gradually reduce the dynamics from *f* to *p* |
| 5 | *Adagio* | Time | Perform the score with 72 bpm |
| 6 | *Largo* | | Perform the score with 50 bpm |
| 7 | *Rit.* | | Gradually reduce the speed from 60 - 50 bpm |
| 8 | *Accel.* | | Gradually increase the speed from 60 - 72 bpm |

- We have complied a new musical shape evaluation dataset[1], including 4116 high-quality piano recordings in 28 classes of musical shapes;
- We propose a S-ResNN method[2] to evaluate the musical shape in the piano pieces for promoting piano education.

II. DATASET DESCRIPTION

Inspired by [2][22], we select a well-established educational book on piano finger practice to cover the possible correlation of finger strength and construct a comprehensive experiment. The exercises composed by both Hanon [23] and Schmitt [24] are two highly admitted textbooks. As Schmitt has more various pitch patterns than Hanon, the latter is highly controversial [25]. Thus, we introduce a new musical shape evaluation dataset here, which has collected 147 music pieces (83 polyphony, 20 scales, 12 arpeggios, and 32 staccato) from Schmitt [24].

There are plenty of musical shapes in terms of various time and dynamics patterns [8]. To validate our concept, we consider the normal musical shape as a speed of 60 beat per minute (bpm) with normal dynamics. A single time pattern in terms of faster and slower speed is performed as 72 bpm and 50 bpm, which are represented as *Adagio* and *Largo* in music theory, respectively. It is worth noticing that the actual performing speed is between 50 (50 bpm * 1 note) - 288 (72 bpm * 4 notes) notes per minute. Stronger and weaker dynamics patterns at 60 bpm are represented as *Forte* and *Piano*, respectively. Taking another 4 time and dynamics patterns into account, 8 basic musical shapes are first determined in Table I, which are then extended to 16 combinations where the corresponding descriptions are shown in Table II. In addition, three supplementary musical shapes, i.e., *Swing*, *Give* and *Take* are also evaluated in our study. *Swing* is the most important feature in Jazz which is a popular music style 20 century [12]. *Give* and *Take* are advanced and delayed movement of time in the music pieces, respectively [8]. Thus, 28 categories of shape (1 normal and 27 specific) are performed on 147 music pieces, resulting in 4116 recordings in WAV format, with a sampling rate of 48 KHz and a period of 7 seconds. In this study, music pieces with normal and 27 musical shapes were played by music trainers and students, respectively. All categories can be clearly separable from listening.

TABLE II. DESCRIPTION OF EXTENDED 16 SHAPES.

| | *Adagio* (72 bpm) | *Largo* (50 bpm) | *Rit.* (60→50 bpm) | *Accel.* (60→72 bpm) |
|---|---|---|---|---|
| *Forte (f)* | *f* + 72 bpm | *f* + 50 bpm | *f* + 60→50 bpm | *f* + 60→72 bpm |
| *Piano (p)* | *p* + 72 bpm | *p* + 50 bpm | *p* + 60→50 bpm | *p* + 60→72 bpm |
| *Cresc. (p→f)* | *p→f* + 72 bpm | *p→f* + 50 bpm | *p→f* + 60→50 bpm | *p→f* + 60→72 bpm |
| *Decresc. (f→p)* | *f→p* + 72 bpm | *f→p* + 50 bpm | *f→p* + 60→50 bpm | *f→p* + 60→72 bpm |

---

[1] https://zenodo.org/record/7225090#.Y0_mCXbMKUk

[2] https://github.com/lixiaoquan731/Musical-Shape-Evaluation

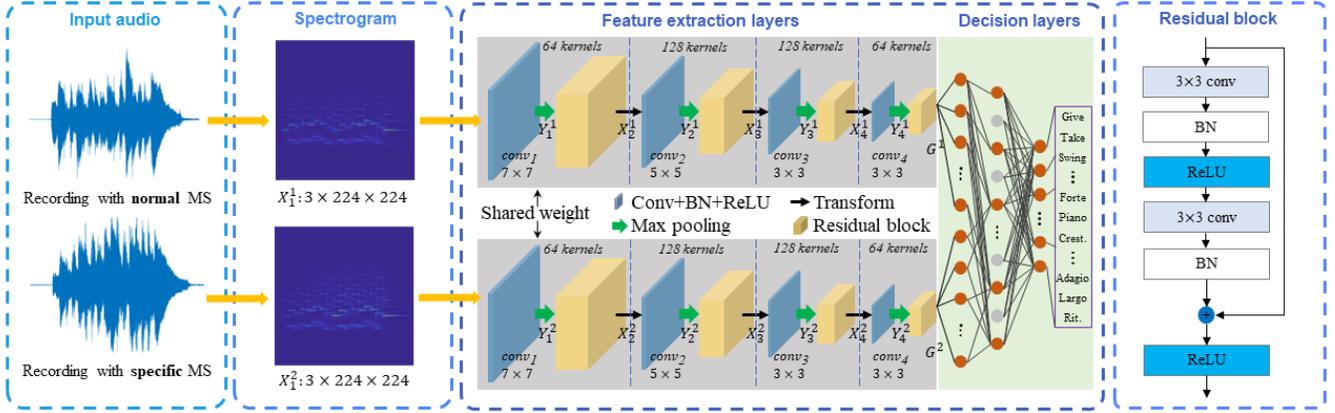

**Fig. 2**. Diagram of S-ResNN. Conv and BN represent the convolutional layer and the batch normalization layer, respectively.

### III. PROPOSED METHOD

Fig. 2 illustrates the architecture of the proposed model as well as essential parameter information. The input of our model is a pair of audio recordings including the music piece with specific musical shape and its corresponding one with normal musical shape. A Constant-Q transform (CQT) [26] is employed to generate the spectrogram for each audio recording. The reason for selecting CQT is because it has been frequently utilised as time-frequency transform in many MIR tasks [27]. Theoretically, any time-frequency transform methods can be fitted into our model.

S-ResNN contains two pyramid-structured branches where the weights are shared. Each branch has four stages comprising a convolutional layer and a residual block. Given an input $X_i^s$ in each stage, where $s \in [1,2]$ is the number of input spectral feature map and $i \in [1,4]$ is the number of stages. The output $Y_i^s$ of each convolutional layer can be expressed as

$$Y_i^s = \sigma(BN(conv_i(X_i^s))) \qquad (1)$$

where $\sigma(\cdot)$ and $BN(\cdot)$ represent the rectified linear unit (ReLU) activation function and batch normalization, respectively. Then the $X_i^l$ will keep updating by

$$X_{i+1}^s = ResBl(MP(Y_i^s)) \qquad (2)$$

where $MP(\cdot)$ denotes a $3 \times 3$ max pooling layer with a stride 2, and $ResBl$ denotes the residual block. The implementation of residual block is shown in Fig. 2, which follows the standard structure in ResNet50 [28]. In each stage, the kernel size of convolutional layer and residual block is the same. The outputs of two branches, $G^1$ and $G^2$, are then concatenated together, followed by a dropout layer $D(\cdot)$ and a flatten layer $F(\cdot)$. To accurately evaluate the musical shapes in the piano pieces, we consider it as a classification task, for which a commonly used loss function, the cross entropy $\mathcal{L}$, is adopted, as defined below.

$$\mathcal{L}_{p,l} = -\frac{1}{t}\sum_{i=1}^{t}(l*log(p) + (1-l)*log(1-p)) \qquad (3)$$

$$p = F(D(concatenate[G^1, G^2]))$$

where t, p and l denote the number of inputs, predicted probability, and the classification label, respectively.

### IV. EXPERIMENTAL SETTING

The proposed S-ResNN is trained on NVIDIA Quadro RTX 6000 with 200 epochs and a batch size of 16. For fast convergence, stochastic gradient descent is selected as the optimizer where the learning rate, momentum and weight decay are set as 1e-3, 0.9 and 0.0005, respectively. The spatial size of the input spectrogram image is set to $3 \times 224 \times 224$.

For quantitative evaluation, three widely used metrics including the precision, recall and F1 score are adopted. Each experiment was repeated 10 times, and averaged results are reported in the Section 4. Within each repetition, training and testing data are randomly selected without overlap. Different training rates ranging from 10% to 70% in each class have been used for training.

To validate the efficacy of our proposed model, comprehensive experiments are carried out where conventional audio content analysis (ACA) methods and deep learning models are used for benchmarking. For the audio methods, some classic MIR techniques including Zero Cross Rate (ZCR), Mel-frequency cepstral coefficients (MFCC), spectral centroid (SpCen), spectral rolloff (SpRf), spectral flux (SpFlux), spectral skewness (SpSkew) and spectral flatness (SpFlat) are used to extract the audio feature followed by a popular classifier i.e., support vector machine (SVM) for the decision making. The models of audio feature extraction and SVM is employed from ACA system [13] and LIBSVM tool [29], respectively. Deep learning models (e.g., VGG16 [16], ResNet50 [17], and DenseNet161 [18]) are employed from Openmmlab's image classification toolbox [30].

### V. RESULTS AND DISCUSSIONS

*A. Comparison with benchmarking methods*

An objective comparison between our method and other benchmarking methods is shown in Fig. 3. As seen, S-ResNN always leads to a higher recall for all training rates. This is due to the fusion of Siamese structure and residual blocks, making

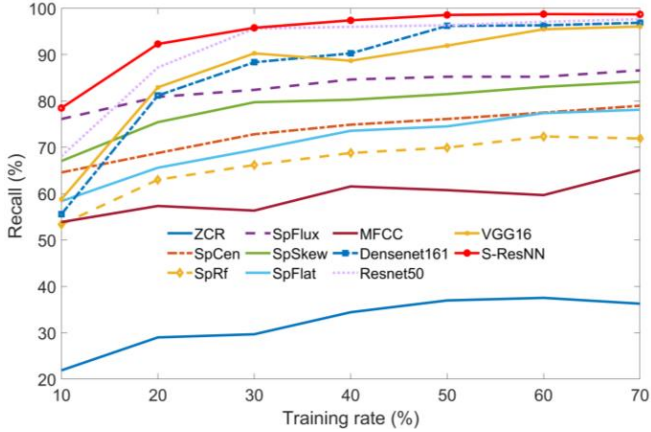

Fig. 3. Comparison of various approaches in dependence of different training rates on Schmitt music recordings.

Table III. Comparison of Classification Performance with Mean Value and Corresponding STD of 7 Training Rates

| Methods | Precision (%) | Recall (%) | F1 score (%) |
|---|---|---|---|
| ZCR | 32.25± 5.29 | 32.75± 5.25 | 32.75± 5.25 |
| MFCC | 59.24± 3.44 | 70.23± 2.56 | 62.20± 3.26 |
| SpCen | 73.37± 4.73 | 74.03± 4.48 | 73.38± 4.72 |
| SpRf | 66.50± 6.16 | 67.23± 5.72 | 66.55± 6.03 |
| SpFulx | 83.01± 3.34 | 83.30± 3.19 | 83.03± 3.31 |
| SpSkew | 78.73± 5.42 | 79.82± 4.67 | 78.87± 5.26 |
| SpFlat | 70.99± 6.55 | 71.90± 6.10 | 71.07± 6.47 |
| VGG16 | 86.39± 13.63 | 87.12± 12.96 | 86.29± 13.86 |
| ResNet50 | 91.13± 9.97 | 91.26± 9.81 | 91.06± 10.11 |
| DenseNet161 | 86.30± 11.95 | 87.04± 10.87 | 86.24± 12.07 |
| S-ResNN | **93.81± 6.60** | **93.98± 6.43** | **93.78± 6.67** |

full use of spectral features to better characterize various musical shapes.

Meanwhile, when the training rate is greater than 20%, the baseline deep learning models consistently produce better results than the ACA methods. This is due to the fact that deep learning models can extract more representative global features than ACA methods with sufficient training data. Thus, the complex time and dynamics patterns in music pieces can be better identified. In addition, the features extracted by ACA methods consist of numerous local temporal and/or spectral features that are insufficient to characterize the musical shapes adequately. The parameter selection (i.e., hop and block size in ZCR and spectral properties, and number of coefficients in MFCC, etc) also affects the classification performance of ACA methods, making them less practicability.

Table III shows the average classification results and standard deviation (STD) of 7 training rates, where it is seen that our S-ResNN is superior to other baseline deep learning models in terms of higher classification accuracy and lower STD. ACA methods have generally lower STD, indicating their stability, but their classification accuracy is much lower than deep learning models under different training rates. On a different point, when the training sample is not sufficient (e.g., 10% training rate as seen in Fig. 3), the baseline deep learning models produce inferior results than some ACA methods such as SpFlux and SpSkew, etc. However, out proposed S-ResNN still yields the best accuracy, which further validate its effectiveness when dealing with limited training data.

### B. Parameter efficiency

Table IV reveals the efficiency of our method and baseline deep learning models. It is observed that our proposed S-ResNN has fewest parameters but needs adequate computation cost. The main reason is that S-ResNN has fewer weighted layer than baseline deep learning models, but the size of fully connected (FC) layer at the end of each branch is $6400 \times 2048$, leading to higher Flops than ResNet50 and DenseNet161. This issue can be potentially solved by reducing the kernel size but increasing the number of convolutional layers with larger stride. With a deeper structure but fewer spatial size of convolutional feature maps, the size of FC layers can be much reduced, and the discriminative information of musical shapes can be well extracted. Thus, a much better balance between effectiveness and efficiency can be achieved.

**Table IV.** Comparison of proposed method and baseline deep learning models on efficiency

| Method | VGG16 | ResNet50 | DenseNet161 | S-ResNN |
|---|---|---|---|---|
| Params (M) | 138.36 | 25.56 | 28.68 | 14.88 |
| Flops (G) | 15.5 | 4.12 | 7.82 | 9.78 |

### C. Generalization experiment

To further validate the robustness and reliability of our proposed model, we have conducted a generalization experiment where 10 music pieces selected from Hanon were performed, resulting 250 music recordings with 25 classes of musical shapes (excluding *Swing*, *Give* and *Take*). Then all models trained on Schmitt music pieces will be directly tested on Hanon music pieces. Comparison of various models using different training rates is shown in Fig. 4. As seen, the proposed S-ResNN has much better generalization ability than other benchmark methods though it produces comparable accuracy to SpFlux in 10% training rate. This actually motivates us to improve the few-shot learning ability of the model by combining data augmentation, attention mechanisms and meta-learning strategies in the future.

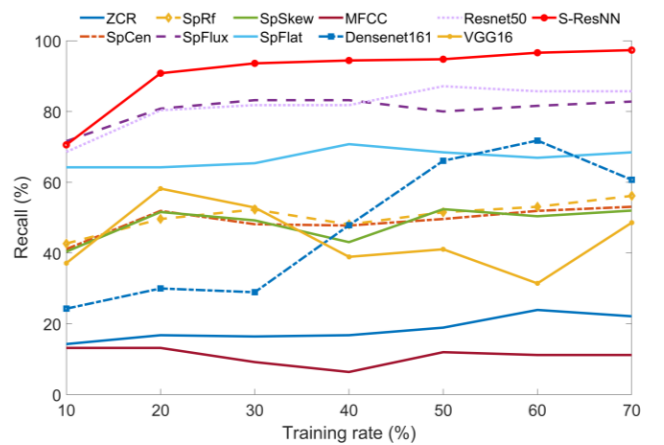

Fig. 4. Comparison of various approaches in dependence of different training rates on Hanon music recordings.

## VI. CONCLUSION

In this paper, we bring a new insight of musical shape evaluation into MPA. Musical shape evaluation is a bridge to link the human perception with music's intrinsic properties, which addresses the shortage of existing MPA framework. A new architecture S-ResNN is proposed for musical shape evaluation, where a new dataset is also built as an extra outcome. Comprehensive experiments have shown our method outperforms not only conventional benchmarking approaches but also several deep learning models. In future work, we plan to further improve and validate our model on wider application scenarios, where the current dataset will be extended to include increased categories of music scores and musical shapes. Furthermore, some open-source tools such as MusicXML and MusPy [31] can be used to adjust the time and dynamics to various levels that are sometimes hardly performed manually.

## REFERENCES

[1] A. Lerch, C. Arthur, A. Pati and S. Gururani, "An interdisciplinary review of music performance analysis," *arXiv Preprint arXiv:2104.09018,* 2021.

[2] H. Kim, P. Ramoneda, M. Miron and X. Serra, "An overview of automatic piano performance assessment within the music education context," in *3rd International Special Session on Computer Supported Music Education,* 2022, p. 456-74.

[3] ABRSM, (2022, Oct.) "ABRSM marking criteria," [Online]. Available:https://gb.abrsm.org/media/12045/markingcriteriaall.pdf.

[4] H. Schreiber, F. Zalkow and M. Müller, "Modeling and estimating local tempo: A case study on chopin's mazurkas." *In Proceedings of the 21th International Society for Music Information Retrieval Conference (ISMIR)*, pp. 773-779, 2020.

[5] A. Vidwans, S. Gururani, C. Wu, V. Subramanian, R. V. Swaminathan and A. Lerch, "Objective descriptors for the assessment of student music performances," in *Audio Engineering Society Conference: 2017 AES International Conference on Semantic Audio,* 2017.

[6] K. A. Pati, S. Gururani and A. Lerch, "Assessment of student music performances using deep neural networks," *Applied Sciences,* vol. 8, *(4),* pp. 507, 2018.

[7] P. Parmar, J. Reddy and B. Morris, "Piano skills assessment," in *2021 IEEE 23rd International Workshop on Multimedia Signal Processing (MMSP),* 2021, pp. 1-5.

[8] D. Leech-Wilkinson, "Musical shape and feeling," *Music and Shape,* vol. 3, pp. 359, 2017.

[9] S. K. Langer, "Philosophy in a New Key: A Study in the Symbolism of Reason, Rite, and Art*," Harvard University Press*, 2009.

[10] Z. Eitan and R. Y. Granot, "How music moves:: Musical parameters and listeners images of motion," *Music Perception,* vol. 23, *(3),* pp. 221-248, 2006.

[11] M. B. Küssner and D. Leech-Wilkinson, "Investigating the influence of musical training on cross-modal correspondences and sensorimotor skills in a real-time drawing paradigm," *Psychology of Music,* vol. 42, *(3),* pp. 448-469, 2014.

[12] A. Lerch, C. Arthur, A. Pati and S. Gururani, "An Interdisciplinary Review of Music Performance Analysis," *Transactions of the International Society for Music Information Retrieval,* vol. 3, *(1),* 2020.

[13] A. Lerch, "An Introduction to Audio Content Analysis: Applications in Signal Processing and Music Informatics,*" Wiley-IEEE Press,* 2012.

[14] K. Kosta, R. Ramírez, O. F. Bandtlow and E. Chew, "Mapping between dynamic markings and performed loudness: a machine learning approach," *Journal of Mathematics and Music,* vol. 10, *(2),* pp. 149-172, 2016.

[15] K. Kosta, "Computational Modelling and Quantitative Analysis of Dynamics in Performed Music," PhD thesis, Queen Mary University, London, UK, 2017.

[16] K. Simonyan and A. Zisserman, "Very deep convolutional networks for large-scale image recognition," *arXiv Preprint arXiv:1409.1556,* 2014.

[17] K. He, X. Zhang, S. Ren and J. Sun, "Deep residual learning for image recognition," in *Proceedings of the IEEE Conference on Computer Vision and Pattern Recognition,* 2016, pp. 770-778.

[18] G. Huang, Z. Liu, L. Van Der Maaten and K. Q. Weinberger, "Densely connected convolutional networks," in *Proceedings of the IEEE Conference on Computer Vision and Pattern Recognition,* 2017, pp. 4700-4708.

[19] G. Koch, R. Zemel and R. Salakhutdinov, "Siamese neural networks for one-shot image recognition," in *ICML Deep Learning Workshop,* 2015, vol. 3, (1).

[20] S. Jadon, "An overview of deep learning architectures in few-shot learning domain," *arXiv Preprint arXiv:2008.06365,* 2020.

[21] W. Rawat and Z. Wang, "Deep convolutional neural networks for image classification: A comprehensive review," *Neural Computation,* vol. 29, *(9),* pp. 2352-2449, 2017.

[22] P. Ramoneda, D. Jeong, E. Nakamura, X. Serra and M. Miron, "Automatic piano fingering from partially annotated scores using autoregressive neural networks," in *Proceedings of the 30th ACM International Conference on Multimedia,* 2022, pp. 6502-6510.

[23] C. L. Hanon, *The Virtuoso Pianist: In Sixty Exercises for the Piano.* G. Schirmer, 1 Nov 1986.

[24] A. Schmitt, *Preparatory Exercises: For the Piano.* G. Schirmer, 1 Nov 1986.

[25] C. C. Chang, "Fundamentals of Piano Practice,*" Citeseer,* 2016.

[26] C. Schörkhuber and A. Klapuri, "Constant-Q transform toolbox for music processing," in *7th Sound and Music Computing Conference, Barcelona, Spain,* 2010, pp. 3-64.

[27] X. Li, Y. Yan, J. Soraghan, Z. Wang and J. Ren, "A music cognition–guided framework for multi-pitch estimation," *Cognitive Computation,* pp. 1-13, 2022.

[28] K. He, X. Zhang, S. Ren and J. Sun, "Deep residual learning for image recognition," in *Proceedings of the IEEE Conference on Computer Vision and Pattern Recognition,* 2016, pp. 770-778.

[29] C. Chang and C. Lin, "LIBSVM: a library for support vector machines," *ACM Transactions on Intelligent Systems and Technology (TIST),* vol. 2, *(3),* pp. 1-27, 2011.

[30] M. Contributors, (2022, Oct.) "Openmmlab's image classification toolbox and benchmark," [Online]. Available: https://Github.Com/Open-Mmlab/Mmclassification.

[31] H. Dong, K. Chen, J. McAuley and T. Berg-Kirkpatrick, "MusPy: A toolkit for symbolic music generation," *In Proceedings of the 21th International Society for Music Information Retrieval Conference(ISMIR),* pp. 101–108., 2020.